\documentclass[runningheads]{style/llncs}

\pdfoutput=1

\newif\ifanonymous

\usepackage[T1]{fontenc}

\usepackage{graphicx}
\usepackage[list=true, font=small, labelfont=bf, labelformat=brace, position=top]{subcaption}

\usepackage{tabularx}
\usepackage{makecell}
\usepackage{multirow}

\usepackage{float}
\usepackage{enumitem}
\usepackage{footmisc}
\usepackage{url}

\usepackage{algorithm}
\usepackage[noend]{algpseudocode} 
\usepackage{amsmath,amsfonts}
\usepackage{bm} 

\setlist[description]{leftmargin=\parindent,}

\newcommand{\Break}{\State \textbf{break}}

\DeclareMathOperator{\argmin}{argmin}

\DeclareMathOperator{\ANNS}{ANNS}
\DeclareMathOperator{\KNN}{KNN}

\DeclareMathOperator{\AND}{\overline{\delta_N}}
\DeclareMathOperator{\DEG}{DEG}

\DeclareMathOperator{\opt}{opt}

\DeclareMathOperator{\extend}{ext}

\DeclareMathOperator{\va}{\large \text{v}_A}
\DeclareMathOperator{\vb}{\large \text{v}_B}
\DeclareMathOperator{\vc}{\large \text{v}_C}
\DeclareMathOperator{\vd}{\large \text{v}_D}

\newcommand\recallAtVar[1]{\operatorname{recall@}\hspace{-0.2em}{#1}}

\ifanonymous
    \newcommand{\repourl}{https://github.com/anonymous}
\else
    \newcommand{\repourl}{https://github.com/Visual-Computing/DynamicExplorationGraph}
\fi

\makeatletter
\newcommand\preprintfootnote[1]{%
  \begingroup
  \renewcommand\thefootnote{}%
  \footnotetext{#1}%
  \addtocounter{footnote}{-1}%
  \endgroup
}
\makeatother

\begin{document}
\title{Dynamic Exploration Graph: A Novel Approach for Efficient Nearest Neighbor Search  \\in
Evolving Multimedia Datasets}
\titlerunning{Dynamic Exploration Graph}
%

\ifanonymous
    \author{anonymous}
\else

\author{
    Nico Hezel\orcidID{0000-0002-3957-4672} 
    \and Kai Uwe Barthel\orcidID{0000-0001-6309-572X} 
    \and Bruno Schilling\orcidID{0009-0006-7021-7311} 
    \and Konstantin Schall\orcidID{0000-0003-3548-0537} 
    \and Klaus Jung\orcidID{0000-0002-3600-6848}
}
\authorrunning{N. Hezel et al.}
%
\institute{
    Visual Computing Group, HTW Berlin, 12459 Berlin, Germany
    \email{\{hezel,barthel,bruno.schilling,konstantin.schall,klaus.jung\}@htw-berlin.de}\\
    \url{https://visual-computing.com/}
}
\fi

\maketitle
\preprintfootnote{\scriptsize \copyright\ The Author(s), under exclusive licence to Springer Nature Singapore Pte Ltd. 2025. Final version: \url{https://doi.org/10.1007/978-981-96-2054-8_25}}
\begin{abstract}
Approximate Nearest Neighbor Search (ANNS) represents a fundamental problem in various applications (image-search, recommendation systems). While graph-based algorithms have demonstrated a good balance between search accuracy and time, handling dynamic datasets, where data points are continuously added or removed, remains a challenge. This paper introduces the Dynamic Exploration Graph (DEG), an extension of the  continuous refining Exploration Graph, which retains high search efficiency for static dataset while adding essential support for dynamic data. At the core of the DEG design are two key innovations: a novel vertex deletion algorithm which guarantees graph connectivity and a data distribution-agnostic method for graph expansion. Through these mechanisms, the DEG maintains a balanced and well-connected structure, even under continuous data alterations. Empirical experiments in both streaming and online scenarios demonstrate the superior performance of the DEG, surpassing existing dynamic graph algorithms in terms of construction time and search efficiency. Although optimized for dynamic datasets, the DEG delivers results as good as current state-of-the-art approaches for static dataset, underscoring its broad applicability.

\end{abstract}

\section{Introduction}

Nearest Neighbor Search finds the closest data points to a query using a distance function on high-dimensional feature vectors. These vectors, often produced by deep learning models, represent multimedia data like images, videos, and audio \cite{Babenko2016,Radford2021}. As the number of data points or feature dimensions increases, linear search becomes computationally expensive, making Approximate Nearest Neighbor Search (ANNS) a more viable solution. It uses data compression or auxiliary structures to skip a substantial portions of the dataset, sacrificing minimal accuracy \cite{Peng2023,Qiang2023} for speed. Proximity graphs provide the best trade-off between accuracy and search speed for static datasets \cite{Wang2021Survey,DPG2020}.

Handling dynamic datasets with continuous additions and removals remains a major challenge. While some proximity graph algorithms, such as the Hierarchical Navigable Small World (HNSW) \cite{Malkov2020} and the continuously refining Exploration Graph (crEG) \cite{Hezel2024}, allow incremental construction, they lack a mechanism for vertex removal. Existing solutions often merely mark deleted vertices without removing them. This approach impairs search speed and overall graph efficiency, as "deleted" vertices must still be considered during the search process. 

This highlights the need for dynamic graph algorithms capable of removing data points without compromising performance. Recent research has focused on algorithms for evolving datasets \cite{Zeng2024,Aguerrebere2024}, mainly in two key scenarios: streaming and online. In streaming scenarios, the index maintains a fixed size, resembling a sliding window over a continuous data stream. Each new data point replaces an old one. This predictability allows simpler deletion mechanisms, as new additions can repair suboptimal connections left by deletions. As a result, algorithms designed for streaming data exhibit low latency when adding or removing data. In contrast, online scenarios are less predictable. The number of data points in the index can vary widely, with no certainty about the frequency or volume of changes.This unpredictability demands dynamic memory management and robust deletion algorithms to remove large portions of the graph while preserving connectivity and search efficiency. Developing an algorithm capable of managing all three data types (static, streaming and online) remains a major challenge.

\section{Contribution}

In this paper, we propose the Dynamic Exploration Graph (DEG), building upon and extending the continuous refining Exploration Graph. Our main contributions include:

\begin{itemize}[label=\textbullet]

\item \textbf{Novel vertex deletion mechanism:} This work introduces a novel graph reduction mechanism following the core principles of crEG. Rather than simply marking vertices as "deleted", it ensures immediate vertex removal while actively repairing the graph to maintain connectivity, regularity and avoid performance degradation.

\item \textbf{Data distribution-agnostic graph expansion:} A new graph expansion method, achieving search efficiency comparable to the extend-and-refine methods of crEG while eliminating the need for a hyper-parameter tailored to the data distribution. This approach offers a significant advantage in dynamic scenarios where the data distribution may change over time.

\item \textbf{Empirical validation of State-of-the-Art performance:} Extensive empirical experiments, encompassing static, streaming and online scenarios, demonstrate the superiority of the DEG over existing graph algorithms. The results highlight the advantages of the DEG: reduced construction time and improved search efficiency. 
\end{itemize}

\section{Related Work}

Graph-based methods are widely used for efficient approximate nearest neighbor search (ANNS) due to their favorable balance of search accuracy and speed. These approaches typically approximate classical proximity graphs, such as k-Nearest Neighbor Graphs (kNNG) \cite{Paredes2005}, Relative Neighborhood Graphs (RNG) \cite{Toussaint1980}, Delaunay Graphs (DG) \cite{Delaunay1933}, and Monotonic Relative Neighborhood Graphs (MRNG) \cite{Cong2019}, as constructing them directly is computationally intensive. The Hierarchical Navigable Small World (HNSW) \cite{Malkov2020} algorithm is one of the most prominent, approximating RNG and DG. A recent alternative, the continuously refining Exploration Graph (crEG) \cite{Hezel2024}, delivers state-of-the-art performance by constructing an exploration graph with properties of DG, RNG, and MRNG through incremental construction and optional edge optimization. These and several other static graph algorithms provide a foundation for understanding proximity graph properties \cite{Wang2021Survey,Oguri2024} and approximation techniques \cite{Jianyang2023,Chen2023}. However, they often struggle to account for dynamic datasets.
\\
\\
\textbf{Streaming Datasets:} Streaming data, with its continuous and high-volume nature, introduces challenges for ANNS, including maintaining accuracy, managing ingestion latency, and adapting to changing data distributions. Benchmarks like CANDY \cite{Zeng2024} evaluate ANNS performance under these conditions, focusing on ingestion latency and adaptability \cite{Aguerrebere2024,Baranchuk2023}. In cases where the index size must remain fixed or only recent data in a data stream should be considered, the sliding window model is applied. In SWINN \cite{Mastelini2024}, data deletion and insertion are combined into an update operation. 
\\
\\
\textbf{Online Datasets:} Online datasets experience less frequent updates than streaming data but can see greater fluctuations in the number of data points. Algorithms in this domain must manage the deletion of large graph portions without sacrificing integrity or search efficiency. HNSW \cite{Malkov2020} and crEG \cite{Hezel2024} are incremental graphs and can be adapted to handle dynamic updates by marking deleted vertices, though this can degrade their performance over time, as shown in Section \ref{sec:experiments}. FreshDiskANN \cite{Singh2021} offers an alternative strategy by using a two-layer architecture. A streaming layer marks data for deletion, while a batch layer periodically merges these changes into the index once approximately 10\% of the vertices have been marked and updates a large number of edges. The HDR+ Tree \cite{Ukey2023} follows a similar strategy, but also introduce a delayed and efficient batch update, to reduce redundant computations and processing all changes in small batches (expansion and reduction). Updates are treated as a combination of deletions and insertions, replacing old entries with new ones. Additionally vertices with similar neighbors are placed in the same batch to reduce the number of edge updates. IPMG \cite{Zhaozhuo2022} and OLGraph \cite{Zhao2020} demonstrate how to integrate changes immediately in a directed graph. Unfortunately, both approaches require a reverse graph, which doubles the storage overhead of the edges and none of them guarantees connectivity or reachablity.

\section{Graph Properties for Dynamic Data}

This section examines existing graph-based approaches and their suitability for dynamic datasets in interactive search and exploration systems.

\begin{itemize}
    
\item \textbf{Incremental Graph:} An incremental graph supports adding new data points without rebuilding the entire graph, making it ideal for large, continuously growing datasets by avoiding costly reconstructions.

\item \textbf{Fully Dynamic Graph:} Fully dynamic graphs extend incremental graphs by allowing both additions and deletions, essential for online datasets or streaming services.

\item \textbf{Search Reachability:} Approximate nearest neighbor algorithms typically start searches from a specific vertex, assuming a path exists from that vertex to any other point, a property called search reachability.

\item \textbf{Graph Connectivity:} While search reachability assumes paths from a specific starting point, graph connectivity ensures a path exists between any two vertices. This is crucial for exploratory search, where starting points may change based on user interactions or deletions.
\end{itemize}

\noindent The Table \ref{tab:graph_properties} compares various graph-based approaches based on these properties, including whether they offer an open-source implementation. HDR+ Tree, OLGraph, and IPMG are still in preprint with no available implementations. The continuously refining exploration graph (crEG) shows desirable properties for static datasets, such as incremental growth, search reachability, and graph connectivity, making it suitable for exploratory search. Our proposed Dynamic Exploration Graph (DEG) builds on these features and introduces improved graph extension and reduction methods to handle fully dynamic datasets.

\vspace{-0.5em}
\begin{table*}[ht!]
\begin{tabular*}{\textwidth}{l@{\extracolsep{\fill}}*{5}{c}}
\hline
\makecell{\textbf{Algorithm}} & 
\makecell{\textbf{Incremental} \\ \textbf{Graph}} &  
\makecell{\textbf{Fully} \\ \textbf{Dynamic} \\ \textbf{Graph}} & 
\makecell{\textbf{Search} \\ \textbf{Reachability}} &  
\makecell{\textbf{Graph} \\ \textbf{Connectivity}} &  
\makecell{\textbf{Open} \\ \textbf{Source}} \\
\hline
HDR+ Tree & \textbf{Yes} & \textbf{Yes} & No & No & No \\
IPMG & \textbf{Yes} & \textbf{Yes} & No & No & No \\
OLGraph & \textbf{Yes} & \textbf{Yes} & No & No & No \\
HNSW & \textbf{Yes} & No & No & No & \textbf{Yes} \\
SWINN & \textbf{Yes} & \textbf{Yes} & No & No & \textbf{Yes} \\
FreshDiskANN & \textbf{Yes} & \textbf{Yes} & \textbf{Yes} & No & \textbf{Yes} \\
crEG & \textbf{Yes} & No & \textbf{Yes} & \textbf{Yes} & \textbf{Yes} \\
DEG & \textbf{Yes} & \textbf{Yes} & \textbf{Yes} & \textbf{Yes} & \textbf{Yes} \\
\hline
\end{tabular*}
\caption{Comparative analysis of graph properties for various ANNS algorithms.}
\label{tab:graph_properties}
\end{table*}
\vspace{-3.0em}

\section{Dynamic Exploration Graph} \label{sec:dynamicExplorationGraph}
The Dynamic Exploration Graph (DEG) builds on the continuous refining Exploration Graph principles \cite{Hezel2024}. This section reviews key crEG concepts and highlights the properties that enable DEG to efficiently manage evolving datasets.
\\
\\
\textbf{Even Regularity: } Both crEG and DEG are even-regular undirected graphs $G(V,E)$ without loops, where $V$ is the set of vertices and $E$ the set of edges. The regularity, denoted by $d$, is expressed as $\DEG_d$. The smallest instance of $\DEG_d$ is the complete graph $K_{d+1}$ with $d+1$ vertices. The number of edges in a regular undirected graph can be derived from the Handshake Lemma \cite{Euler1736} and is given by $|E| = (|V| \cdot d)/2$. Even regularity is crucial, as any odd-regular undirected graph requires an even number of vertices, which is unattainable if the dataset contains an odd number of data points. A minimum regularity of 4 is required to avoid cycles and isolated vertices.
\\
\\
\textbf{Connectivity: } A key feature of DEG is its guaranteed connectivity, crucial for efficient exploration. This is achieved by maintaining an Eulerian graph structure, where each vertex has an even degree and undirected edges, ensuring an Euler cycle traversing all edges. As a result, at least two disjoint paths exist between any pair of vertices, with no bridges present. This 2-edge connectivity allows the removal of an edge without disconnecting the graph.
\\
\\
\textbf{Edge Quality: } The dynamic nature of DEG necessitates adding and removing vertices, requiring edge modifications to preserve graph properties. A crucial aspect is evaluating the impact of edge swaps on the overall graph quality. While traditional metrics like Graph Quality (GQ) prove insufficient in capturing the effects of minor graph alterations, crEG introduces the Average Neighbor Distance ($\AND$) as a more fine-grained metric using any distance function $\delta(\cdot,\cdot)$.

\begin{equation} \label{eq:averageNeighborDistance}
\AND(U) = \frac{1}{|U|} \sum_{u \in U} \frac{1}{d} \hspace{-0.1cm}\sum_{v \in N(G, u)} \delta(u, v)
\end{equation}
\\
This metric quantifies the average distance between connected vertices, with lower values indicating a higher degree of similarity among neighbors $N(G, u)$. By storing distances in the weights of the edges $( w_{u, v} = \AND(u, v) )$, computing $\AND(U)$ for any set of vertices $U$ becomes very efficient.

\subsection{Graph Reduction} 
\begin{figure}[h!]
\centering
\includegraphics[trim=2cm 10.5cm 2.5cm 0.5cm,clip=true,width=\columnwidth]{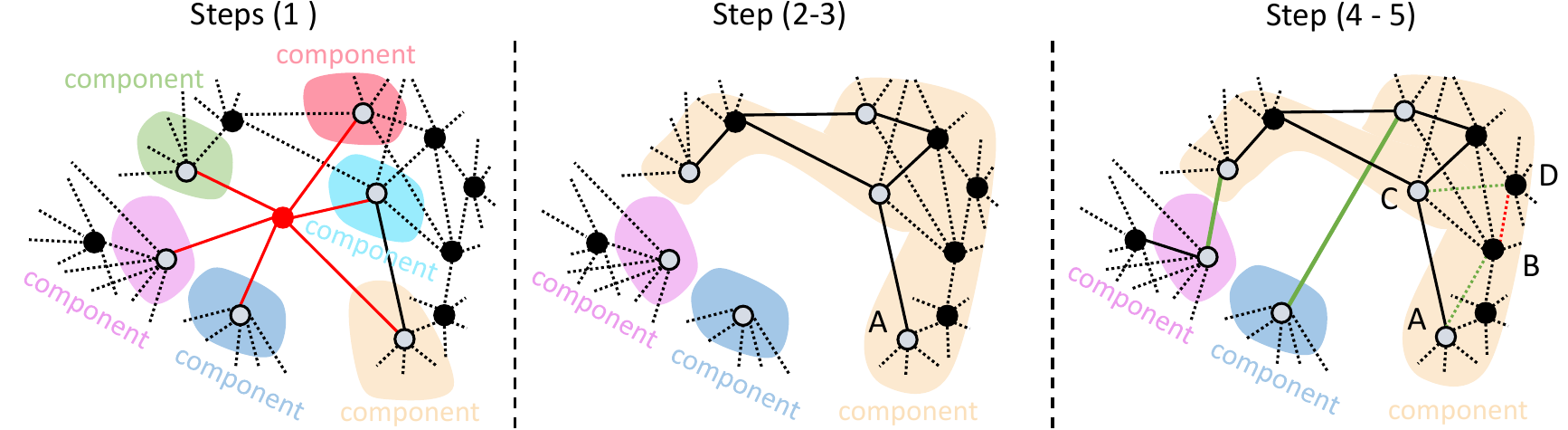}
\vspace{-0.5cm}
\captionsetup{font=small}
\caption{A 2D toy example of a $\DEG_{\bm{6}}$. Left: The red vertex is removed, the gray vertices form there individual a vertex induced subgraphs. Middle: Reachable subgraphs are merged. Right: Sufficient paths (black solid lines) between subgraphs were identified to connect the remaining subgraphs with new edges (green solid lines).} \label{fig:graph_reduction}
\end{figure}

Removing a vertex $v$ from the Dynamic Exploration Graph requires isolating the vertex and reconnecting its neighbors $V_i = N(G, v)$ to preserve the graph's connectivity. Due to the Eulerian property of DEG, every neighbor $u \in V_i$ has an alternative path to $v$ through another vertex in $V_i$ \cite{Toida1974}. Consequently, after removing $v$, each vertex in $V_i$ can still reach at least one other vertex in $V_i$ via a remaining path. However, this does not ensure full connectivity among all vertices in $V_i$, potentially leading to fragmentation into disconnected components. Since the neighbors in $V_i$ were connected to $v$ based on their similarity, they are likely to be close to each other within the graph. This proximity makes it feasible to restore connectivity by exploring their neighborhoods and reconnecting any resulting components, as illustrated in Figure \ref{fig:graph_reduction}.

\vspace{-0.5cm}

\noindent \textbf{Connectivity Restoration:} After vertex $v$ and its edges are removed using Algorithm \ref{alg:reduceGraph}, each of its former neighbors $u \in V_i$ is missing an edge and the graph might be disconnected. The restoration of its connectivity can be described in 5 steps. All special cases and further details are provided in Algorithm \ref{alg:restoreGraph}.

\begin{enumerate}
\item \textbf{Induced Subgraph Formation:} A vertex-induced subgraph $G'_u$ is created for each neighbor $u \in V_i$, initially containing only $u$.

\item \textbf{Neighborhood Exploration:} A breadth-first search (BFS) is initiated simultaneously from all vertices in $V_i$, with each subgraph $G'_u$ expanding incrementally as new reachable vertices are added.

\item \textbf{Subgraph Merging:} The BFS continues until a vertex from another subgraph is encountered, at which point the two subgraphs are merged, as they are now mutually reachable.

\item \textbf{Circular Reconnection:} Once only $d / 2 - 1$ subgraphs remain, edges are added between the vertices in $V_i$ to form a circular arrangement, ensuring each vertex in $V_i$ can reach every other either directly or through the merged subgraphs.

\item \textbf{Restoration of Regularity:} Finally, the algorithm introduces additional edges as needed to ensure all vertices in $V_i$ are even regular again.
\end{enumerate}

\vspace{-0.5cm}
\begin{algorithm}[h!]\small
    \caption{ReduceGraph($G, d, v$)}
    \label{alg:reduceGraph}
    \begin{algorithmic}[1]
    \Require current graph $G$, even degree $d \in \mathbb{N}$, vertex $v$ to remove
    \Ensure remove $v$ from $G$
	
    \State $V_i \gets N(G,v)$ \Comment{current neighbors of v}
	\State remove $v$ and its edges from $G$ \Comment{remove the vertex and its edges}  
    \If{$|V(G)| \leq d$} \Comment{small DEGs are already fully connected}
        \State \Return 
    \EndIf
    
    \State restoreGraph($G,V_i$,true) \Comment{reconnect the possible graph components}
    \end{algorithmic}    
\end{algorithm}

\begin{algorithm}[h!]\small
    \caption{RestoreGraph($G, V_i$, optimize)}
    \label{alg:restoreGraph}
    \begin{algorithmic}[1]
    \Require graph $G$, set of vertices $V_i$ missing an edge, flags to optimize new edges
    \Ensure reconnect the vertices in $V_i$ to restore connectivity of $G$

    \State $R \gets$ EmptyMap \Comment{associates a vertex with a set of reachable vertices}
    \ForAll {$v \in V_i$} \Comment{Step1: each involved vertex can reach itself}  
        \State $R(v) \gets \{v\}$
    \EndFor  

    \State $c \gets 0$  \Comment{merge counter}  
    \State $V_q \gets$ FIFO queue initialized with $V_i$ in any order\Comment{vertices to check}
    \While {$c < |V_i|/2-1$} \Comment{Step2: find other subgraphs}
        \State $v \gets $ pop element from $V_q$ 
        \State $S \gets N(G,v) \cup \{n\}$ \Comment{neighbors of $v$} 
        \State $S \gets S \cup (\bigcup_{s \in S} R(s))$ \Comment{reachable vertices of $v$}
        \State $c \gets c + |S \cap V_i| - |R(v) \cap V_i|$ \Comment{count new reachable vertices}
        \ForAll {$s \in S$} \Comment{every vertex in $S$ can reach all vertices in $S$}
            \State $R(s) \gets S$
        \EndFor  
        \State add $N(G,v)$ to $V_q$ in any order \Comment{check neighbors of $v$ later}
    \EndWhile

    \State remove all entries in $R$ where its key is not in $V_i$
    \State $E_i \gets \emptyset$ \Comment{improvable new edges}    
    \State $L \gets$ list of unique reachable sets from the values of $R$ in any order
    \While {$|L| > 1$} \Comment{Step3: reconnect subgraphs}
        \State $V_f \gets L.\text{first} \cup V_i$ \Comment{first subgraph} 
        \State $V_l \gets L.\text{last} \cup V_i$ \Comment{last subgraph} 
        \State $u,v \gets \argmin_{u \in V_f} \argmin_{v \in V_l} \delta(u,v)$ \Comment{find good match} 
        \State add edge $(u,v)$ to $G$ and add to $E_i$ \Comment{connect the subgraphs} 
        \State $L.\text{first} \gets V_f \cup V_l$ \Comment{merge subgraphs information} 
        \State remove $L$.last \Comment{remove last subgraph from $L$} 
        \State $V_i \gets V_i \setminus \{u, v\}$ \Comment{update the list of vertices missing an edge}    
    \EndWhile
      
    \ForAll {$v \in V_i$} \Comment{Step4: reconnect the remaining involved vertices} 
        \State $u \gets \argmin_{u \in V_i} \delta(v,u)$ where $N(G,v) \cap \{u\} = \emptyset$
        \If{$\{u\} \neq \emptyset$}
            \State add edge $(v,u)$ to $G$ and add to $E_i$
            \State $V_i \gets V_i \setminus \{v, u\}$
        \EndIf
    \EndFor

    \ForAll {$\va \in V_i$} \Comment{Step5: handle edge cases}
        \State $V_n \gets \bigcup_{i \in N(G,\va)} N(G,i)$ \Comment{2-hop neighborhood of $\va$} 
        \State $\vb \gets$ any $\vb \in V_n$ where $\va \neq \vb$ AND $N(G,\va) \cap \{\vb\} = \emptyset$
        \ForAll {$\vc \in V_i$}
            \State $\vd \gets$ any $\vd \in N(G,\vb)$ where $\vd \neq \va$ AND $\vd \neq \vc$ AND $N(G,\vc) \cap \{\vd\} = \emptyset$

            \State remove edge $(\vb,\vd)$ from $G$
            \State add edge $(\va,\vb)$ to $G$ and add to $E_i$
            \State add edge $(\vc,\vd)$ to $G$ and add to $E_i$
            \State $V_i \gets V_i \setminus \{\va, \vc\}$
            \Break
        \EndFor
    \EndFor
    
    \If{optimize}
        \State OptimizeEdge(G,u,v,...) for $e=(u,v) \in E_i$\Comment{optionally improve edges}
    \EndIf
    \end{algorithmic}
\end{algorithm}

\newpage

\subsection{Graph Extension} 

We propose a novel graph extension method, which initially connects a new vertex to its best neighbor candidates and then restores graph regularity and connectivity using the repair mechanism from the deletion process. This method matches the efficiency of crEG's extend and refine techniques while eliminating the need for a hyper-parameter tuned to the data distribution. 

When incorporating a new vertex $v$ into the graph, DEG connects $v$ to the most closest candidates $b$ identified through a RangeSearch \cite{Iwasaki2010}. The edge $(v, b)$ is preferably added if it approximates the Monotonic Relative Neighborhood Graph (MRNG), ensuring a well-structured connection. If $b$ already has $d$ edges, the longest edge $(b, n)$ is removed to create space for the new edge $(v, b)$. In cases where $b$ has previously lost an edge, $v$ can connect directly to $b$. Removing an edge may disrupt the graph's regularity and, potentially, its connectivity. To address this, DEG uses the graph restoration algorithm at the end of the extension process. This algorithm restores regularity and connectivity by reconnecting vertices that lost edges, including the newly added vertex $v$, ensuring the graph is fully connected again. For details, refer to Algorithm \ref{alg:extendGraph}. The algorithm calls "RestoreGraph" with "optimize = false", indicating the restore method should not try to optimize edges added during restoration. 

\begin{algorithm}[h!]\small
    \caption{ExtendGraph($G, d, v, S, k_{\extend}, \varepsilon_{\extend}$)}
    \label{alg:extendGraph}
    \begin{algorithmic}[1]
    \Require current graph $G$, graph degree $d \in \mathbb{N}$, new vertex $v$, set of seed vertices $S$, number of search results $k_{\extend} \in \mathbb{N}$ with $k_{\extend} \geq d$, search range factor $\varepsilon_{\extend} \in \mathbb{R}^+$
    \Ensure connect $v$ to the closest vertices in $G$

    \State $S \gets$ RangeSearch($G, S, v, k_{\extend}, \varepsilon_{\extend}$)
	\Comment{find $k$ vertices similar to $v$}
    \State \textit{skipMRNG} $\gets$ false \Comment{two phases: with and w/o MRNG checks}

    \State $U \gets \emptyset$ \Comment{new neighbors of $v$}
    \State $V_i \gets \{v\}$ \Comment{set of vertices potentially missing an edges}
    \While {$|U| < d - 1$} \Comment{stop if enough neighbors have been found}
         \State $B \gets S \setminus U$ \Comment{not yet connected candidates}
         \While {$|U| < d - 1$ AND $B \neq \emptyset$} \Comment{check all candidates} 
            \State $b \gets \argmin_{x \in B} \delta(x, v)$ \Comment{closest candidate to $v$ in $B$}
            \State $B \gets B \setminus \{b\}$ \Comment{ignore candidate next iteration}
            
            \If{\textit{skipMRNG} $=$ true OR CheckMRNG($G, v, b$)}
                \If{$N(G,b) = d$} \Comment{if $b$ is not yet missing an edge}
                    \State $N \gets N(G,b) \setminus V_i$ \Comment{find another candidate connected to $b$}
                    \State $n \gets$ vertex in $N$ with longest edge to $b$
                    \State remove edge $(n,b)$                     
                    \State $V_i \gets V_i \cup \{n\}$ \Comment{remember $n$ is missing an edges}
                \EndIf
                \State add edge $(v,b)$ 
                \State $U \gets U \cup \{b\}$ \Comment{remember $b$ is connect to $v$}                
            \EndIf
         \EndWhile
         \State \textit{skipMRNG} $\gets$ true \Comment{second phase without MRNG checks}
    \EndWhile

    \State $V_i \gets \{\forall i \in V_i, |N(G,i)| < d\}$ \Comment{update the list of vertices missing an edge}
    \State RestoreGraph($G, V_i$, false) \Comment{reconnect the possible graph components}
    \end{algorithmic}    
\end{algorithm}

\noindent \textbf{Approximating DG and MRNG:} The graph extension algorithm ensures DEG approximates both a Delaunay Graph (DG) and a Relative Neighborhood Graph (RNG). This approximation enhances the efficiency of range searches within the graph. DEG employs Algorithm \ref{alg:checkMRNG} to identify potential neighbors satisfying the edge properties of a Monotonic Relative Neighborhood Graph (MRNG). Edges conforming to the criteria are preferably integrated into DEG, improving its navigational efficiency.
\vspace{-1.0em}
\begin{algorithm}\small
    \caption{CheckMRNG($G, v1, v2$)} 
    \label{alg:checkMRNG}
    \begin{algorithmic}[1]
    \Require Graph $G$, vertex $v1 \in V$ of $G$, vertex $v2 \in V$ of $G$
    \Ensure An edge between $v1$ and $v2$ is MRNG-conformant
    \ForAll {$u \in N(G, v1) \cap N(G, v2)$}
        \If{$\delta(v1, v2) > max(w_{v1, u}, w_{v2, u})$)} 
            \State \Return false
        \EndIf
    \EndFor
    \State \Return true
    \end{algorithmic}  
\end{algorithm}
\vspace{-1.5em}

\section{Experiments}
\label{sec:experiments}
To ensure a comprehensive comparison, we evaluated our \textbf{DEG}\footnote{\repourl} implementation against several state-of-the-art graph-based algorithms: \textbf{HNSW}\footnote{https://github.com/nmslib/hnswlib/tree/master/hnswlib}, \textbf{DiskANN}\footnote{https://github.com/microsoft/DiskANN}, and \textbf{SWINN}\footnote{https://github.com/online-ml/river}, focusing on index build time and the trade-off between search speed and recall rate. The serial scan results from \textbf{FAISS}\footnote{https://github.com/facebookresearch/faiss} served as a baseline. Unfortunately, the implementation of SWINN is not very efficient for datasets of this size. As noted in the original paper, SWINN achieves only a 6x speedup over serial scan with 5000 data points. In our tests, this advantage decreases further, rendering SWINN slower than the serial scan. All experiments were conducted on a Ryzen 5600G at 4.3GHz with 64GB of DDR4 memory at 3200MHz. Graph indices were kept in memory to avoid I/O bottlenecks. The multimedia datasets used are frequently cited in related literature \cite{Wang2021Survey,Cong2021,DPG2020,Chen2023} and are listed in Table \ref{tab:datasets}. The local intrinsic dimension (LID) \cite{Facco2017} for each dataset was computed to better reflect the complexity of its data distribution.

\begin{table}[h]
    \centering
    \begin{tabularx}{\linewidth} { 
   >{\arraybackslash}p{2.5cm} 
   >{\arraybackslash}X  
   >{\arraybackslash}X 
   >{\arraybackslash}X 
   >{\arraybackslash}X 
   >{\arraybackslash}X  }
        \hline
        \textbf{Dataset} & \textbf{Dims} & \mbox{\textbf{\# Base}} & \mbox{\textbf{\# Query}} & \textbf{TopK} & \textbf{LID}\\ 
        \hline
        SIFT1M \cite{Jegou2011} & 128 & 1,000,000 & 10,000 & 100 & 9.2\\
        Deep1M \cite{Babenko2016} & 96 & 1,000,000 & 10,000 & 100 & 15.5\\    
        GloVe \cite{Pennington2014} & 100 & 1,183,514 & 10,000 & 100 & 21.7\\ 
        \hline
    \end{tabularx} 
    \vspace{0.5em}
    \caption{Details of the used datasets: dimensionality of the feature vectors, number of data points and queries, and local intrinsic dimension of the dataset.}
    \label{tab:datasets}
    \vspace{-2.0em}
\end{table}

\subsection{Evaluation metrics and hyper-parameters} 
\label{sec:evaluationMetrics}

We evaluate the search performance of the different algorithms based on their search speed (queries per second (QPS)) in relation to their $\recallAtVar{k}$. The average $\recallAtVar{k}$ is formally defined as:

\begin{equation}
\recallAtVar{k} = \frac{1}{|Q|} \sum_{q \in Q} \frac{\lvert \ANNS(P, q) \cap \KNN(P, q) \rvert}{k}
\label{eq:recallAtK}
\end{equation}

\noindent where $\ANNS$ is the answer set returned by the algorithm, and $\KNN$ is the ground-truth set of the given query. During testing, the $k$ nearest neighbors (e.g., 100-NN) for all queries are retrieved, and the average $\recallAtVar{k}$ and QPS are plotted.
\\
\\
Since real-world applications prioritize search speed in high-recall areas, we focused on achieving a recall above 95\% and adjusted the hyper-parameters accordingly. The dynamic exploration graph consistently used the same parameters across all datasets, with each new vertex connected to 30 neighbors. The parameters $k_{\extend} = 60$ and $\varepsilon_{\extend} = 0.1$ were applied for the range search when identifying suitable candidates during graph extension. For vertex deletion, the OptimizeEdge method from crEG where used following the original paper's settings: $k_{\opt} = 30$, $\varepsilon_{\opt} = 0.001$, and $i_{\opt} = 5$. The graph construction parameters for other methods were sourced from their respective papers or from this survey \cite{Wang2021Survey}, and are detailed on our project page.

\subsection{Search quality in static, streaming and dynamic datasets} \label{sec:dynamicSearchExperiments}

The following section evaluates the approaches across various scenarios. Although DiskANN primarily focuses on SSD-stored indices \cite{Subramanya2019}, it also supports in-memory indices. Its extension, FreshDiskANN \cite{Singh2021}, includes a graph reduction feature. Additionally, we used HNSW's mark-vertex-as-deleted function to observe how a typically static graph behaves in dynamic datasets. We tested three different scenarios using the datasets:

\begin{itemize}[label=\textbullet]
\item \textbf{T1:} Build the graph with the first half of the dataset to establish the baseline index. Similar to a static dataset scenario.

\item \textbf{T2:} Construct the graph using the entire dataset, then remove all vertices from the second half of the dataset from the index. Similar to a dynamic dataset scenario.

\item \textbf{T3:} Build the graph with the second half of the dataset and iteratively remove a vertex from the second half while simultaneously add a vertex from the first half. Similar to a streaming dataset scenario.
\end{itemize}

\begin{figure*}[ht!]
    \begin{subfigure}{0.32\textwidth}
        \includegraphics[trim=2.0cm 5.5cm 2.0cm 7.5cm,clip=true,width=\textwidth]{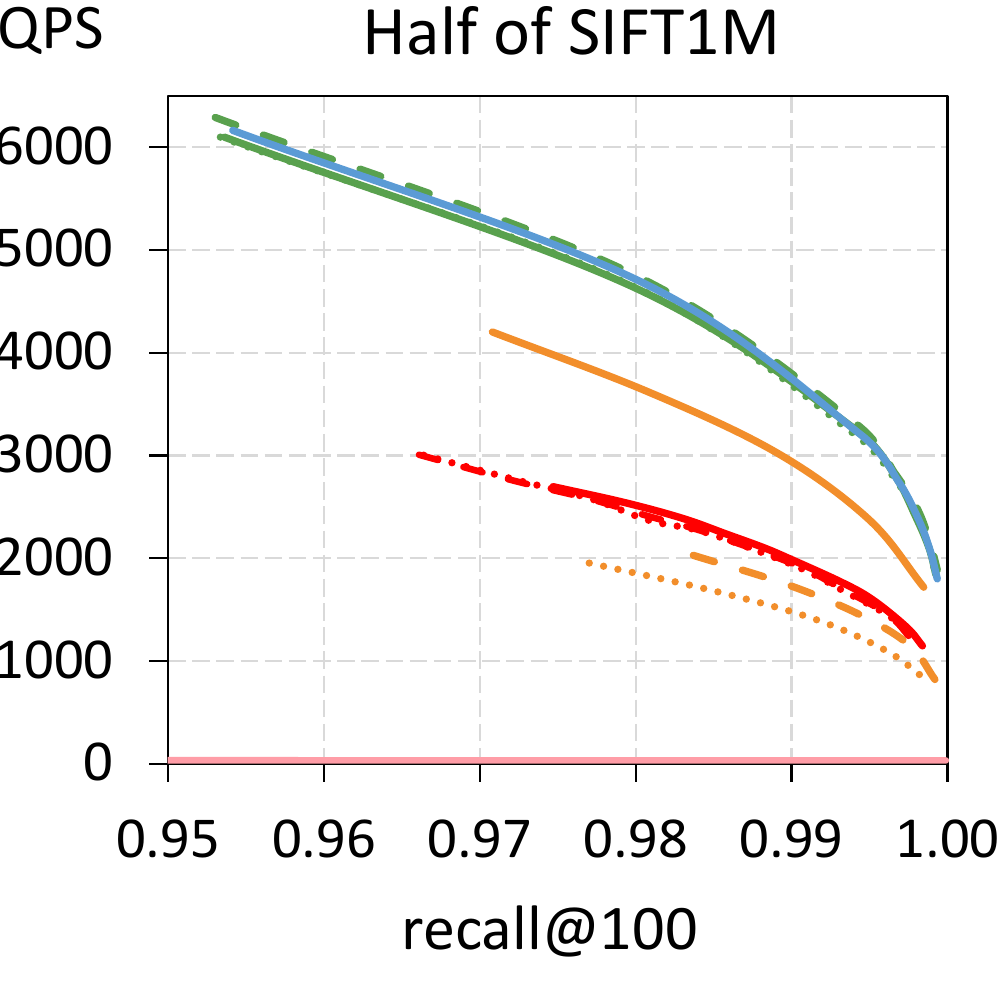}
    \end{subfigure}
    \hfill
    \begin{subfigure}{0.32\textwidth}
        \includegraphics[trim=2.0cm 5.5cm 2.0cm 7.5cm,clip=true,width=\textwidth]{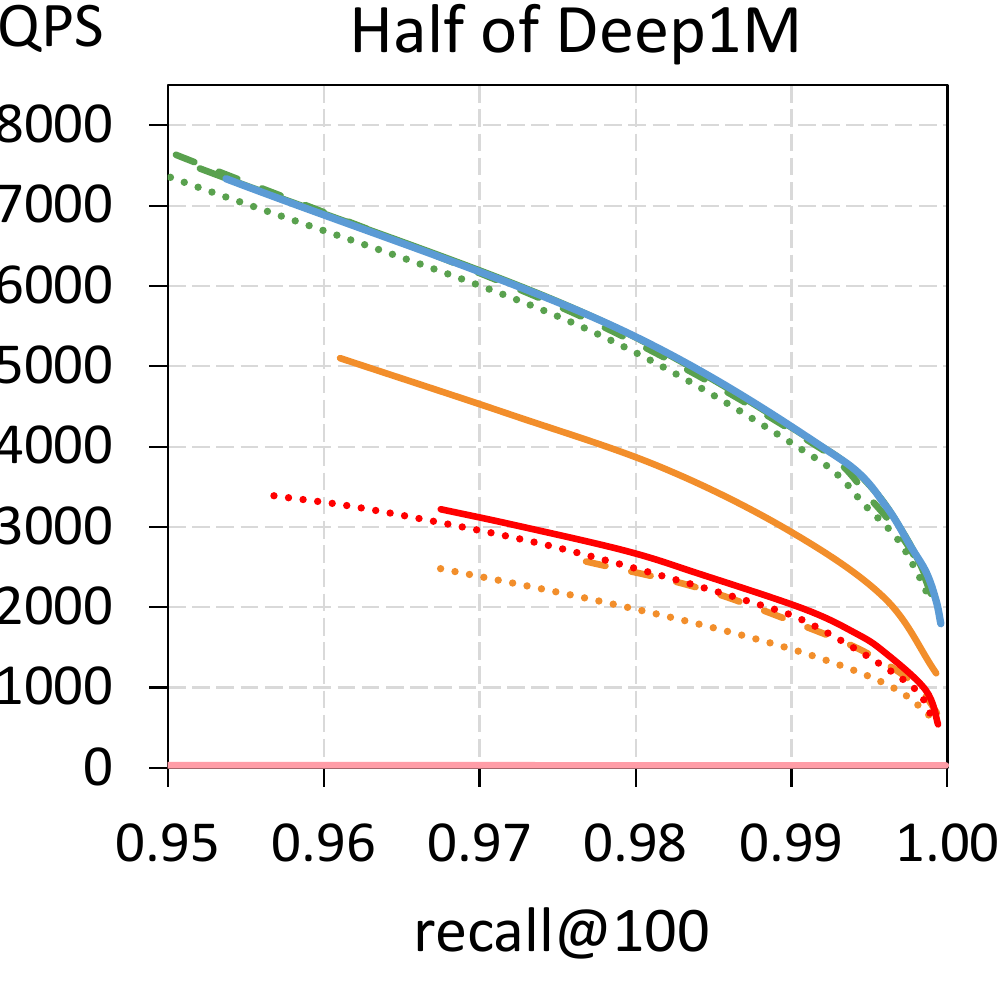}
    \end{subfigure}
    \hfill
    \begin{subfigure}{0.32\textwidth}
        \includegraphics[trim=1.9cm 5.5cm 2.0cm 7.4cm,clip=true,width=\textwidth]{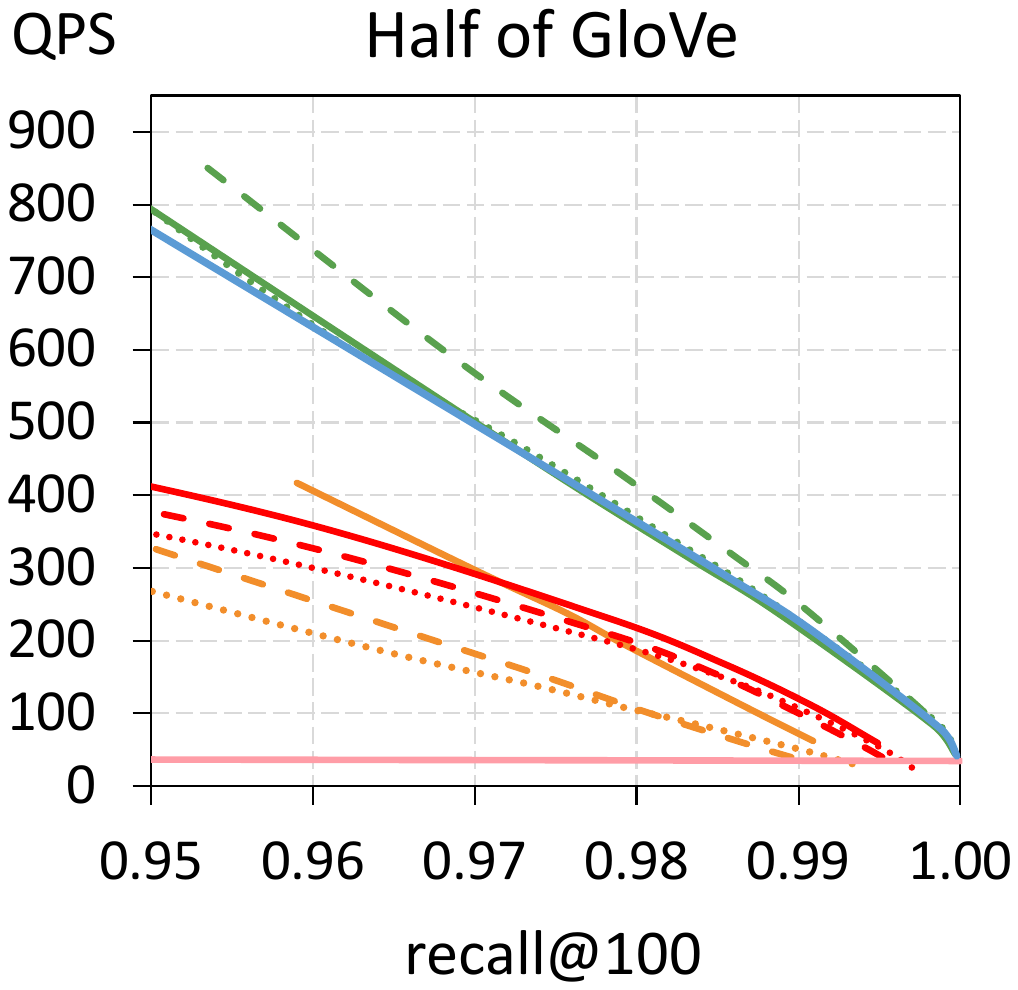}
    \end{subfigure}
    \begin{subfigure}{0.95\textwidth}
        \includegraphics[trim=1.75cm 13.60cm 2cm 15.0cm,clip=true,width=\textwidth]{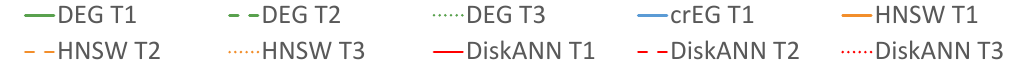}
    \end{subfigure}
    \caption{The number of queries per second (QPS) in relation to recall@100 for approximate nearest neighbor search was tested for different graphs, datasets and use cases (top right is better). }
    \label{fig:search_performance}
\end{figure*}

\noindent Figure \ref{fig:search_performance} presents the results of the three scenarios. DEG shows the highest search efficiency, further improving when vertices are removed due to additional edge optimization during deletion. In the static scenario (T1), DEG even outperforms HNSW and is on par with crEG. For dynamic datasets, HNSW's performance is nearly halved when every second vertex is marked for removal. DiskANN initially marks vertices for deletion and then removes them during consolidation. We follow the suggestions of the authors and consolidate after marking 10\% of vertices. DiskANN's performance is consistent across all scenarios and quite slow compared to the other approaches. Table \ref{tab:indexing_speed} shows the corresponding indexing times, with DEG being the fastest across all tasks and datasets. crEG can be faster without edge optimization. However, we adhere to the parameter settings of the original paper, adjusting only the number of edge optimizations, reducing them by half to account for using only half the dataset size.

\begin{table*}[ht!]
    \centering
    \begin{tabularx}{\textwidth} { |
   >{\centering\arraybackslash}l  |
   >{\centering\arraybackslash}X  
   >{\centering\arraybackslash}X  
   >{\centering\arraybackslash}X  |
   >{\centering\arraybackslash}X  
   >{\centering\arraybackslash}X  
   >{\centering\arraybackslash}X  |
   >{\centering\arraybackslash}X   
   >{\centering\arraybackslash}X  
   >{\centering\arraybackslash}X  |}
        \hline
        \multirow{2}{*}{

        \parbox{1.7cm}
        { \centering
        \ \\[-0.7ex]
        \textbf{Dataset}
        \ \\[2.6ex]
        \textbf{Algorithm}
        }
        
        } &
        \multicolumn{3}{c|}{\textbf{SIFT1M}} &
        \multicolumn{3}{c|}{\textbf{Deep1M}} &
        \multicolumn{3}{c|}{\textbf{GloVe}} \\ 
        \cline{2-10}
        &
        $\bm{T1}$ \textbf{(min)} & 
        $\bm{T2}$ \textbf{(min)} & 
        $\bm{T3}$ \textbf{(min)} & 
        $\bm{T1}$ \textbf{(min)} & 
        $\bm{T2}$ \textbf{(min)} & 
        $\bm{T3}$ \textbf{(min)} & 
        $\bm{T1}$ \textbf{(min)} & 
        $\bm{T2}$ \textbf{(min)} & 
        $\bm{T3}$ \textbf{(min)} \\ 
        \hline
        HNSW & 10.2 & 23.7 & 28.5 & 6.8 & 16.1 & 19.3 & 16.4 & 36.6 & 44.7 \\   
        DiskANN & 7.1 & 22.0 & 21.3 & 6.7 & 18.1 & 17.9 & 11.2 & 35.1 & 41.3 \\
        SWINN & 114.9 & - & 372.4 & 108.2 & - & 336.8 & 157.6 & - & 481.3 \\   
        crEG & 5.3 & - & - & 4.1 & - & - & 12.4 & - & - \\ 
        DEG & \textbf{3.3} & \textbf{11.0} & \textbf{10.0} & \textbf{3.1} & \textbf{11.0} & \textbf{9.9} & \textbf{7.8} & \textbf{30.0} & \textbf{27.3} \\ 
        \hline
    \end{tabularx}
    \vspace{0.1cm}
    \caption{Single threaded indexing time of different approaches for the three tasks.}
    \label{tab:indexing_speed}
    \vspace{-1.0cm}
\end{table*}

\section{Ablation Study} \label{sec:scalabilityAndComplexity}
\begin{figure*}[ht!]
    \begin{subfigure}{0.32\textwidth}
        \includegraphics[trim=5cm 2cm 5.4cm 2.2cm,clip=true,width=\columnwidth]{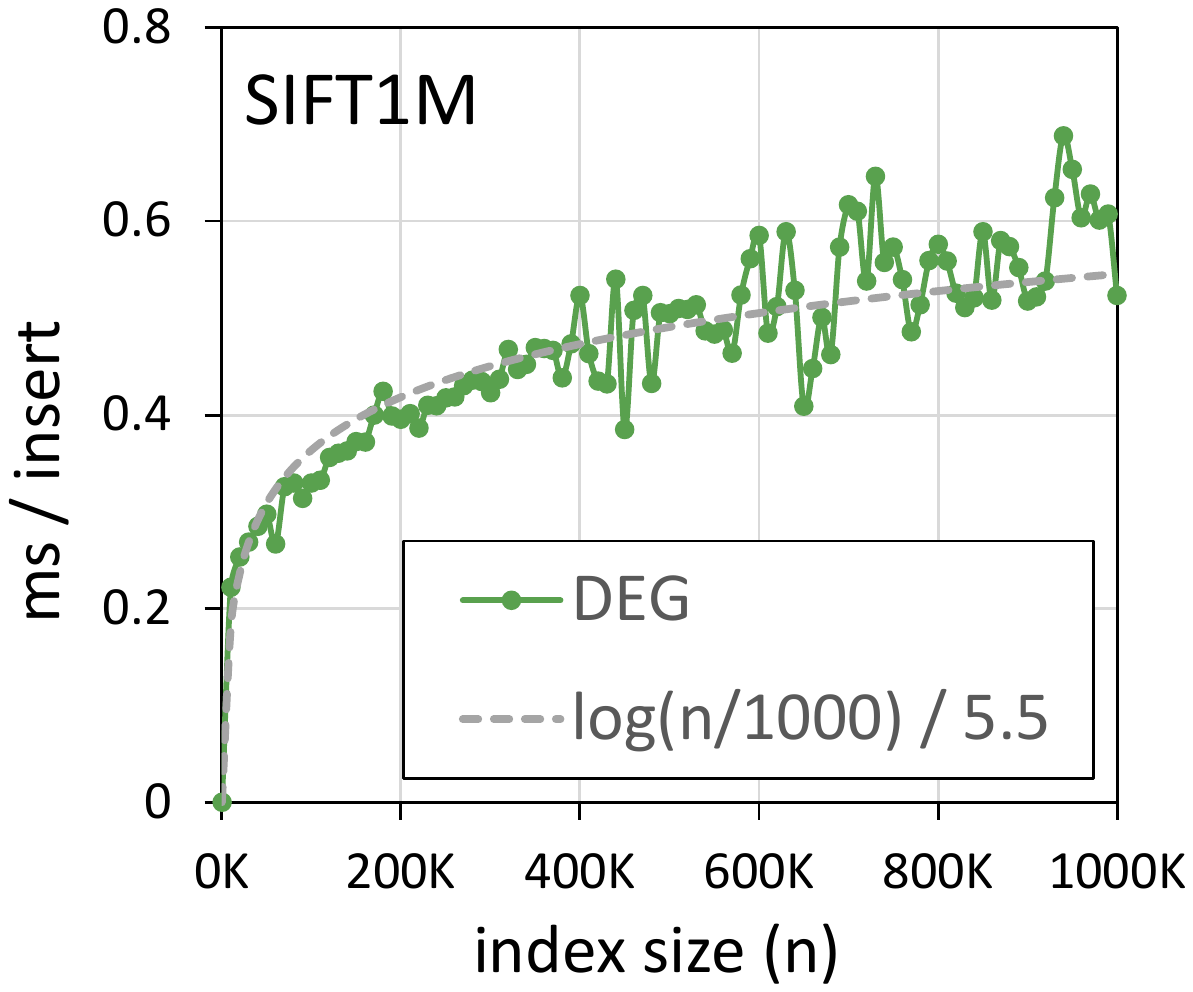}
    \end{subfigure}
    \hfill
    \begin{subfigure}{0.32\textwidth}
        \includegraphics[trim=5cm 2cm 5.4cm 2.2cm,clip=true,width=\columnwidth]{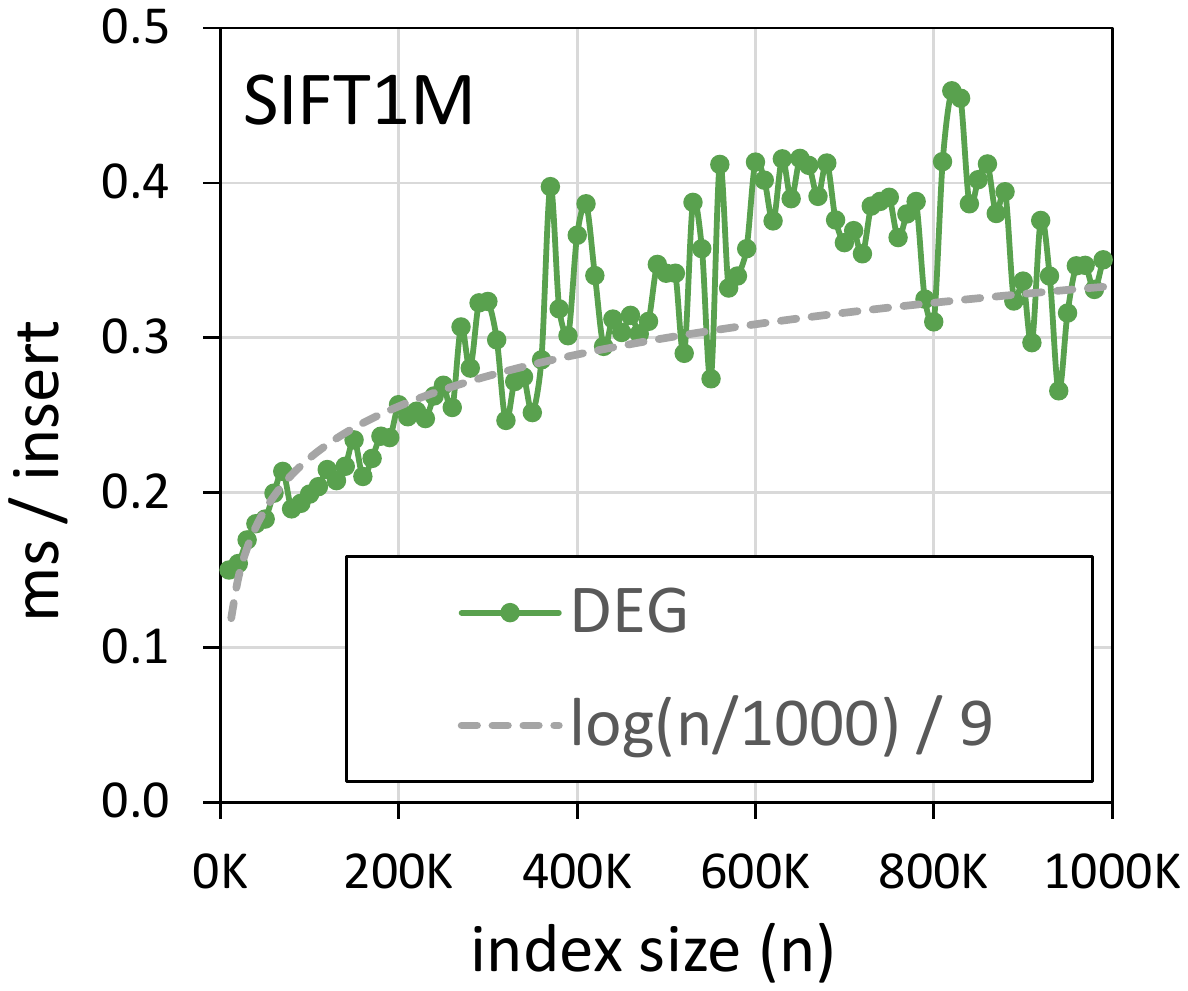}
    \end{subfigure}
    \hfill
    \begin{subfigure}{0.32\textwidth}
        \includegraphics[trim=1.9cm 6.8cm 2.2cm 8.5cm,clip=true,width=\textwidth]{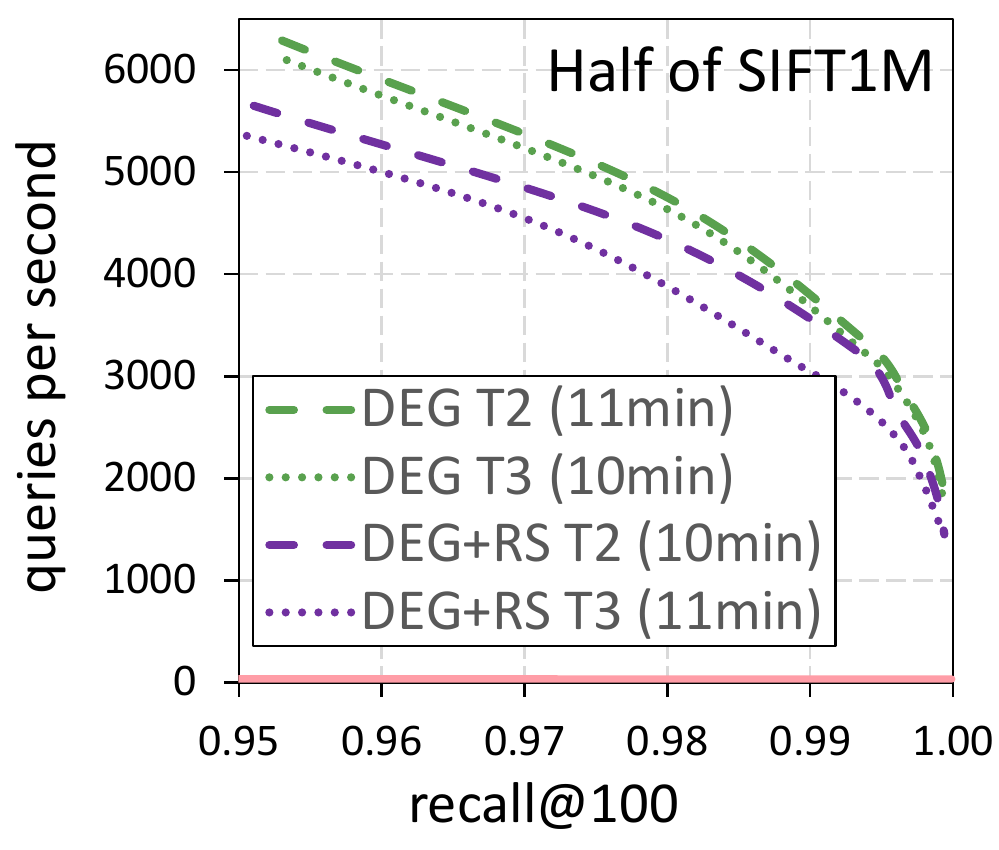}
    \end{subfigure}
    \caption{Influence of index size on indexing speed (left) and reduction speed (middle). The right plot demonstrates the performance of graph reduction when BFS is replaced with RangeSearch.}   \label{fig:build_time_and_reduction_tests}
\end{figure*}

To better assess the scalability of our algorithms, we incrementally constructed DEG using the complete SIFT1M dataset, followed by the gradual removal of all vertices from the graph. We documented the average processing time for every 10,000 modifications, which is presented in the plots (left and middle) in Figure \ref{fig:build_time_and_reduction_tests}. To extrapolate this data and predict the time required for larger datasets, various curve-fitting functions were applied. Notably, the time complexity appears logarithmic and is comparable to that of crEG for graph construction. This aligns with observations from other graphs approximating classical proximity graphs like DG and RNG, as noted in \cite{Wang2021Survey}.

\paragraph{BFS vs RangeSearch for graph reduction: } During the restoration of DEG using Algorithm \ref{alg:restoreGraph}, we employ BFS to identify subgraphs reachable from other vertices missing an edge (involved vertices). In a well-constructed proximity graph, most involved vertices tend to be in close proximity, following kGraph's principle, "the neighbors of my neighbors might be my neighbor." This property enables the identification of large subgraphs containing many involved vertices without requiring more complex search methods like RangeSearch. Smaller subgraphs are then easily connected to the larger ones. In the experiment shown in Figure \ref{fig:build_time_and_reduction_tests} on the right, we replaced BFS with RangeSearch and found that RangeSearch, which only determines the reachability between two vertices, performed worse than identifying larger subgraphs using BFS.

\section{Conclusion} 
\label{sec:conclusion}

This paper introduces the Dynamic Exploration Graph (DEG), a novel approach for efficient nearest neighbor search in dynamic multimedia datasets. DEG builds upon the strengths of the continuous refining Exploration Graph (crEG) while introducing a new vertex deletion algorithm and a data distribution-agnostic graph expansion method. Empirical evaluations demonstrate DEG's superior performance in static, streaming, and online scenarios, surpassing existing algorithms in construction time and search efficiency. DEG's ability to maintain a balanced and well-connected structure, even under continuous data alterations, makes it a powerful tool for various applications requiring real-time search capabilities in evolving datasets.

\newpage
%
%
%
\bibliographystyle{splncs04}
\bibliography{references}
\end{document}